\documentclass[fleqn,showpacs,preprintnumbers,amsmath,amssymb, prb, twocolumn]{revtex4-1}

\usepackage{stmaryrd}
\usepackage{txfonts}
\usepackage{amssymb}
\usepackage{mathrsfs}
\usepackage{graphicx}
\usepackage{dcolumn}
\usepackage{bm}
\usepackage{epsfig}
\usepackage[usenames, dvipsnames]{color}
\usepackage{xcolor}

\begin{document}

\title{Vortices with Fractional Flux Quanta in Multi-Band Superconductors}

\author{Zhao Huang\(^{1,2}\) and Xiao Hu\(^{1,2}\)}

\affiliation{\(^{1}\) {International Center for Materials Nanoarchitectonics (WPI-MANA), National Institute for Materials Science, Tsukuba 305-0044, Japan}
\\
\(^{2}\)Graduate School of Pure and Applied Sciences, University of Tsukuba, Tsukuba 305-8571, Japan}

\begin{abstract}
In superconductors with three or more components, time-reversal symmetry may be broken when the inter-component couplings are repulsive, leading to a superconducting state with two-fold degeneracy. When prepared carefully there is a stable domain wall on a constriction which connects two bulks in states with opposite chiralities. Applying on external magnetic field, vortices in different components dissociate with each other, resulting in a ribbon shape distribution of magnetic field at the position of domain wall.
\end{abstract}

\date{\today}

\maketitle

\section{Introduction}
The discoveries of superconductivity in $\textrm{MgB}_2$ \cite{Nagamatsu01} and iron pnictides \cite{Kamihara08} lead to wide interest in multi-band superconductors. Multiple condensates couple each other through interband interactions, which not only increase the critical temperature \cite{Kondo63} but also induce unconventional pairing symmetries \cite{Mazin08,Kuroki08}. In a two-band superconductor, two order parameters have the same and opposite phases for the attractive and repulsive coupling respectively. When there are three or more bands with interband repulsions, it is interesting that a frustrated state where phase differences among order parameters are neither $0$ nor $\pi$ appears, where time-reversal symmetry (TRS) is broken \cite {Agterberg99, Stanev10,Tanaka10,Carlstrom11, Yanagisawa12, Dias11, Hu12, Maiti13, Orlova13, Wilson13, Takahashi14}, leading to interesting phase sensitive phenomena such as massless Leggett mode \cite{Lin12} and asymmetric critical current \cite{Huang14}.

\begin{figure}[t]
\psfig{figure=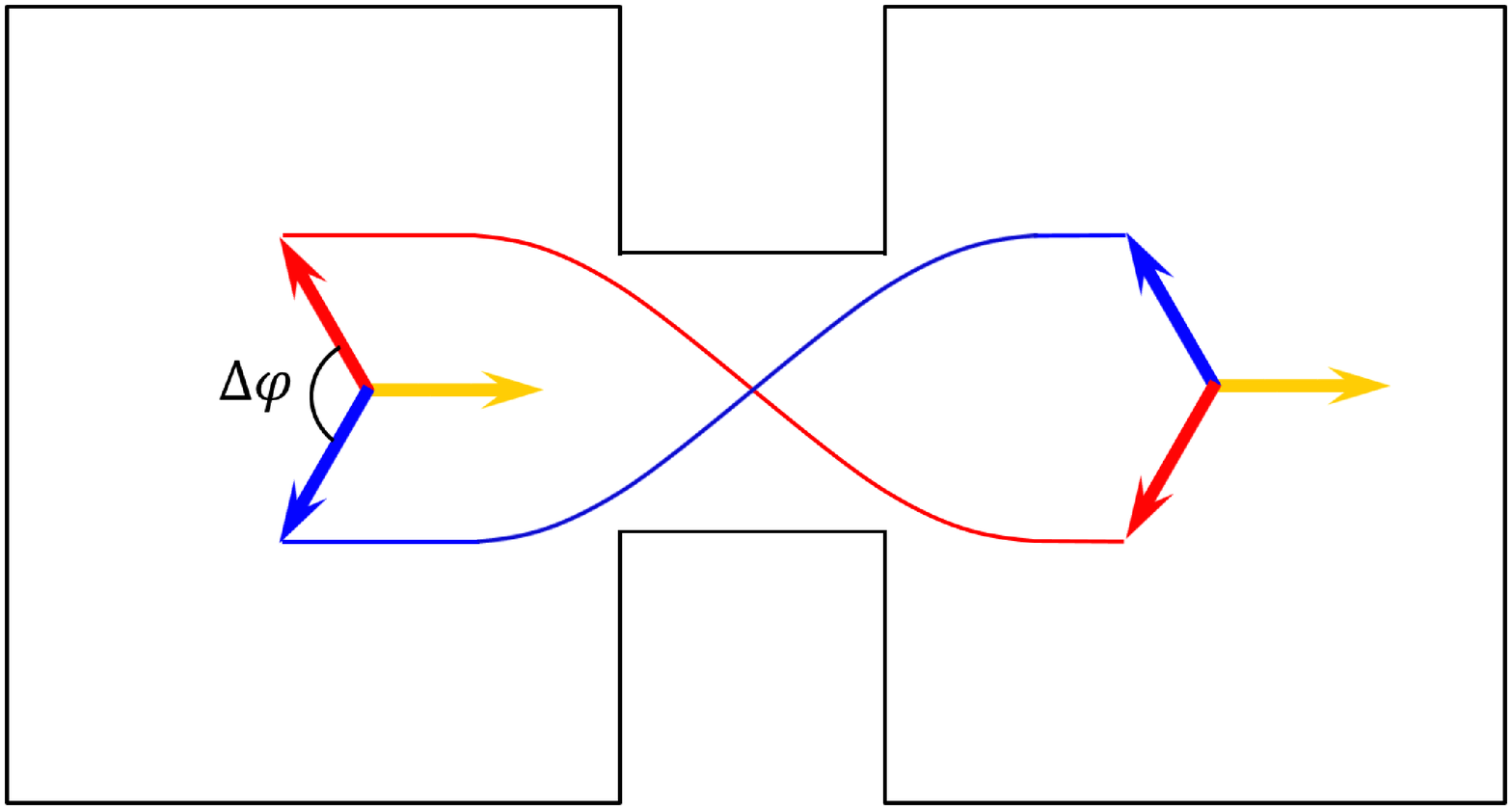,width=\columnwidth} \caption{\label{f1} Schematics of constriction junction between two bulks with distinct TRSB states. A domain wall forms on the constriction where phase difference between red and blue order parameters $\Delta\varphi$ closes and opens again.}
\end{figure}

In a type II superconductor, magnetic field penetrates into sample in terms of vortices associated with tiny magnetic fluxes. In single-band case, a vortex, namely a $2\pi$ phase rotation of the superconducting order parameter, carries quantized magnetic flux $\Phi_0$. Multi-band superconductors are different because multiple condensates simultaneously couple to a common gauge field, and a vortex associated with $2\pi$ phase rotation in one condensate only carries a fraction of $\Phi_0$ \cite{Babaev02,Bluhm06,Tanaka10,Vakaryuk12,Gillis14,Huang15}. When interband phase differences are locked to each other due to strong couplings, phase rotation along a closed path is the same for different condensates. Therefore, vortices in different bands overlap in space, which leads to conventional integer flux quantization. Therefore, in most cases vortices with fractional flux quanta cannot be observed experimentally. The situation is different when a phase separation exists between the two time-reversal symmetry broken (TRSB) superconducting states with opposite chiralities in a sample. On domain walls \cite{Sigrist99,Hu12,Garaud11,LinN12}, interband phase difference develops a phase kink as shown in Fig. \ref{f1}, resulting in different winding numbers in different components in presence of external magnetic field. In the present work we consider a constriction junction where such a domain wall is stabilized due to the small size of junction.

The remaining part of the paper is organized as follows. Time-dependent Ginzburg Landau (TDGL) equations and TRSB states are introduced in Sec. II for three-component superconductors. Then we present simulation results with fractional vortices in Section III, and make discussions and conclusions in Section IV.

\section{Model and numerical technique}
We adopt the Ginzburg-Landau (GL) free energy of three-component superconductors with Josephson-like inter-component couplings \cite{Zhitomirsky04, Gurevich07}
{
\small{
\begin{equation}\label{GLenergy}
\begin{split}
&F=\sum\limits_{j,k=1,2,3}{\left[{\alpha_j(T)}\left|\psi_j\right|^2+\frac{\beta_j}{2}\left|\psi_j\right|^4+\frac{1}{2m_j^*}\left|\left(\frac{\hbar}{i}
\nabla-\frac{e^*}{c}{\bf{A}}\right)\psi_j\right|^2\right]}\\
&-\sum\limits_{j<k}\gamma_{jk}\left(\psi_j^*\psi_k+c.c.\right)+\frac{1}{8\pi}\left(\nabla\times \bf{A}\right)^2,
\end{split}
\end{equation}
}}where $\psi_j=|\psi_j|e^{i\varphi_j}$ is the superconducting order parameter of the $j\textrm{th}$ component. For each component, we have the temperature dependent coefficient $\alpha_j(T)$, which is negative when $T<T_{cj}$, and positive when $T>T_{cj}$ \cite{Gurevich07}. The critical temperature of bulk superconductivity $T_c$ is higher than $T_{cj}$ when we have nonzero inter-component couplings \cite{Kondo63}. Here we consider our system at temperature close to $T_c$, where $\alpha_j$ is positive. $\gamma$ is taken independent of temperature for simplicity. The inter-component coupling is attractive or repulsive when \(\gamma_{jk}\) is positive or negative respectively. Other terms are all conventionally defined \cite{Tinkhambook, Gropp96}.

For the convenience of analysis, we introduce the dimensionless quantities by making the following substitutions
{
\small{
\begin{equation}
\begin{split}
&\psi_j=\psi_{10}(0)\psi_j',\ x=\lambda_1(0)x',\ {\bf{A}}=\lambda_1(0)H_{1c}(0)\sqrt{2}{\bf{A}'},\\
&\gamma_{jk}=-\alpha_1(0)\gamma_{jk}',\ F=\frac{H_{1c}^2(0)}{4\pi}F',\ {\bf{J}}=\frac{e^*\hbar\psi_{10}^2(0)}{m_1^*\xi_1(0)}{\bf{J'}},
\end{split}
\end{equation}
}}where $\bf{J}$ is the current density, $\psi_{10}(0)=\sqrt{-\alpha_1(0)/\beta_1}$, $\lambda_1(0)=\sqrt{-m_1^*c^2\beta_1/4\pi\alpha_1(0)e^{*2}|}$, $\xi_1(0)=\sqrt{-\hbar^2/2m_1^*\alpha_1(0)}$ and $H_{1c}(0)=\sqrt{-4\pi \alpha_1(0)\psi_{10}^2(0)}$. We define $\kappa_1=\lambda_1(0)/\xi_1(0)$ for latter use. The free energy in the dimensionless form is thus given by
{
\small{
\begin{eqnarray}
F'&=&\sum\limits_{j,k=1,2,3}{\left[-\frac{\alpha_j}{\alpha_1(0)}\left|\Psi_j\right|^2+\frac{\beta_j}{2\beta_1}\left|\Psi_j\right|^4+\frac{m_1^*}{m_j^*}\left|\left(\frac{1}{i\kappa_1}
\nabla-{\bf{A}}\right)\Psi_j\right|^2\right]}\nonumber \\
&-&\sum\limits_{j<k}\gamma_{jk}\left(\Psi_j^*\Psi_k+c.c.\right)+\left(\nabla\times \bf{A}\right)^2. \label{eq: a}
\end{eqnarray}
}
}

In the present work, we study superconductors with external magnetic field where the order parameters are inhomogeneous. A direct minimization of GL free energy is difficult in this case. It is convenient to adopt the TDGL approach which describes time evolution processes relaxing the system to a stable state \cite{Gropp96}. The TDGL equations in the zero-electric potential gauge are given by
\begin{eqnarray}\label{TDGL}
\frac{\partial}{\partial t}\psi_j   =   - \frac{{\delta F'}}{{\delta {\psi_j ^*}}},\ \sigma\frac{\partial}{\partial t}{\bf{A}} =   - \frac{1}{2}\frac{{\delta F'}}{{\delta {\bf{A}}}},
\end{eqnarray}
where $\sigma$ is the normal conductivity. For the external boundary, we take the superconductor-insulator boundary condition which means that supercurrent cannot flow in and out of the sample. We can see that at the stable state, the right-hand sides of Eq. (\ref{TDGL}) are zero which give the multi-component GL equations
\begin{eqnarray}
&-\frac{\alpha_j}{\alpha_{1}(0)}\psi_j+\frac{\beta_j}{\beta_1}|\psi_j|^2\psi_j+\frac{m_1^*}{m_j^*}\left(\frac{1}{i\kappa_1}\nabla-\bf{A}\right)^2\psi_j \nonumber&\\
&-\gamma_{jk}\psi_k-\gamma_{jl}\psi_l=0,\\
&{\bf{J_s}}=\nabla\times\nabla\times{\bf{A}}=\sum_{j=1,2,3}\frac{m_1^*}{m_j^*}|\psi_j|^2\left(\frac{1}{\kappa_1}\nabla\varphi_j-\bf{A}\right). \label{GL2}&
\end{eqnarray}

Here we take all components as s-wave for simplicity. When \(\gamma_{12}\gamma_{13}\gamma_{23}>0\), we have solutions with trivial phase differences (0 or \(\pi\)) among order parameters. When \(\gamma_{12}\gamma_{13}\gamma_{23}<0\), a frustrated solution is possible, characterized by nontrivial inter-component phase differences and broken TRS \cite{Hu12}, as shown in Fig \ref{f1}.

\begin{figure}[t]
\psfig{figure=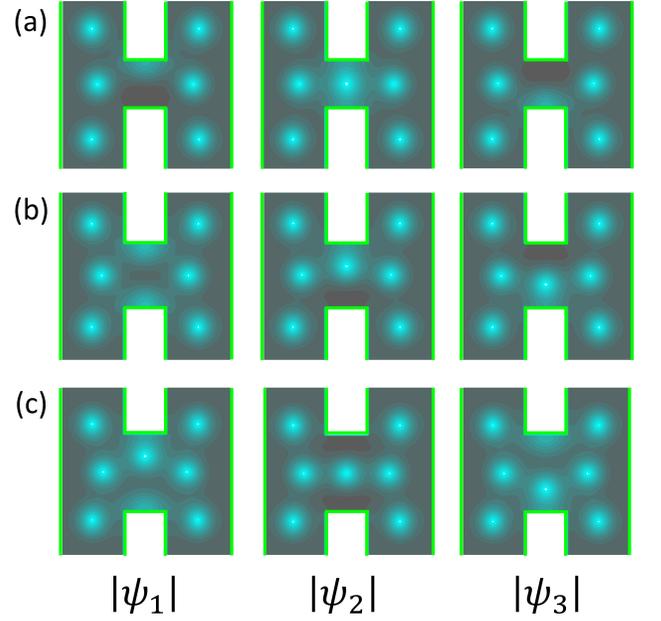,width=8.5cm} \caption{\label{f2} (color online). Simulation results of stable vortex distribution in three components in presence of domain wall on constriction. All panels are contour figures of amplitudes of three order parameters where red, blue and yellow in Fig. \ref{f1} correspond to component-1, 2 and 3 respectively. Area in panel is $10\times10$ which is a part of a larger sample with boundaries shown by green lines. Width of constriction in (a), (b) and (c) is 2.6, 3.6 and 4.4. Parameters are $\alpha=0.2$, $H=0.25$, $\gamma=-0.3$ and $\kappa_1=4.0$. }
\end{figure}

\begin{figure}[t]
\psfig{figure=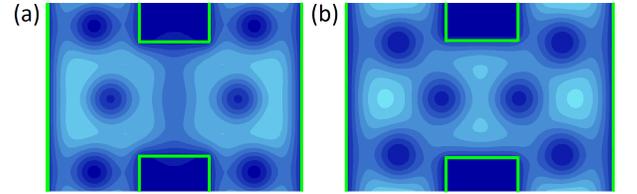,width=8.2cm} \caption{\label{f3} (color online). Magnetic field distribution on constriction when two bulks are at (a) two distinct TRSB states (b) an identical TRSB state. Parameters are the same as Fig. \ref{f2}. Area in panel is $10\times7$ which is a part of a large sample with boundaries shown by green lines. Width of constriction is the same as Fig. 2(c).}
\end{figure}

As an icon we mainly focus here on the isotropic case \((\alpha_1=\alpha_2=\alpha_3\equiv\alpha,\beta_1=\beta_2=\beta_3\equiv\beta,m_1^*=m_2^*=m_3^*\equiv m^*,\gamma_{12}=\gamma_{13}=\gamma_{23}\equiv\gamma<0)\). $T_c$ is determined by $\alpha(T_c)+\gamma=0$. The degenerate TRSB states are $\hat\Psi=|\psi| (1, e^{i2\pi/3}, e^{i4\pi/3})$ and $\hat\Psi^*=|\psi| (1, e^{-i2\pi/3}, e^{-i4\pi/3})$ where $|\psi|=\sqrt{-\alpha-\gamma}$ for $T<T_c$.

In a TDGL simulation, the sample is discretized into a grid of meshes. The mesh size should be carefully selected because a too small size increases the simulation time, which may cause insufficient relaxation, and large sizes may give artificial results. Here, we carefully select the mesh size as $\Delta x=\Delta y=0.1\lambda_1(0)$ when parameters are $\alpha=0.2, \gamma=-0.3$ and $\kappa_1=4.0$. We confirmed that further decreasing the mesh size does not change the results.

\vspace{0mm}
\section{Vortex with fractional flux quanta}
Here we consider a structure of two bulks connected by a constriction as shown in Fig. \ref{f1}. The size of constriction is adjusted to be comparable with the coherence length. We focus on the case with distinct TRSB states ($\hat\Psi$ and $\hat\Psi^*$) in the two bulks, and thus a domain wall forms on the constriction where configuration of three order parameters deforms from the rigid $2\pi/3$ structure. For example, phase difference between blue and red order parameters closes once and opens again, which forms a phase kink as shown in Fig. \ref{f1}.

A laser heat pulse can be used to realize this situation in experiments \cite{Tate89}. At temperature above $T_c$, we irradiate the heat pulse on the constriction. When we cool the sample below $T_c$, the two bulks transform to superconducting states with distinct TRSB states by chance. We then switch off the heat pulse. Superconductivity recovers on the constriction and a domain wall forms. After that we apply the magnetic field to bring vortices into the superconductor.

\begin{figure}[t]
\psfig{figure=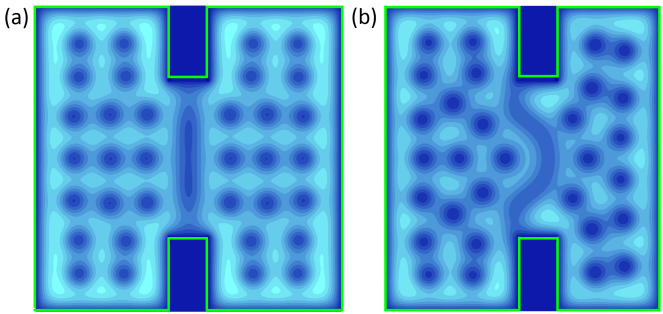,width=8.8cm} \caption{\label{f4} (color online). Magnetic field distribution in presence of vortices with fractional flux quanta for different sample shapes. Parameter are the same as Fig. \ref{f2}. The sample size is $20\times20$ for (a) and $18\times20$ for (b) with boundaries shown by green lines. Width of the constriction is 10.}
\end{figure}

In simulations, two bulks are put at TRSB states with opposite chiralities as the initial condition of TDGL equations. The self-consistent evolution drives the system to a stable state with a domain wall on the constriction. Then we switch on the external magnetic field $H$, and the system evolves again with flux penetrating into the sample from the boundaries and forming vortices inside. A typical vortex configuration near the constriction is shown in Fig. \ref{f2} for the final stable state. It is interesting to find that on the constriction only $\psi_2$ has a $2\pi$ phase winding but $\psi_1$  and $\psi_3$ do not, as seen in Fig. \ref{f2}(a).

We then increase the width of constriction, vortices in component-1 and component-3 start to penetrate the constriction as shown in Figs. \ref{f2}(b) and (c). We can see that vortex cores in different components do not overlap on the constriction, in contrast to those outside the constriction area. When vortices in all components overlap, all order parameters at the core are zero which allow strong magnetic field penetration. While on the constriction, only one order parameter is zero at vortex core, and thus the penetrated field is weak due to the screening from the other two components. These vortices with fractional flux quanta form a ribbon-like distribution of magnetic field as shown in Fig. \ref{f3}(a) and Fig. \ref{f4}. In a symmetric sample, the vortex configuration should be symmetric as seen in Fig. \ref{f4}(a). When the constriction shifts from center as shown in Fig. \ref{f4}(b), the ribbon becomes asymmetric as well. This is different from a previous work \cite{Garaud14}, where irregular ribbon structures were obtained despite of symmetric sample. When the two bulks at the two sides of constriction take a same TRSB state, there is only integer vortices as shown in Fig. \ref{f3}(b). \vspace{3 mm}

\section{Discussions and conclusions}

The vortices with fractional flux quanta discussed in this paper are based on the structure of domain walls in TRSB states. Therefore, the creation and stability of domain wall are important. In the present work, we use a controllable method: create a domain wall with laser heat pulse and stabilize the domain wall by shrinking the its size. In this way, the domain wall can be stable as long as the section of constriction is much smaller than the bulk, even for superconductors with strong interband couplings \cite{Popvich10}.

To summarize, in superconductors with three or more components, a stable domain wall forms on a small junction between two time-reversal symmetry broken states with opposite chiralities. Under external magnetic field, vortices with fractional flux quanta appear on the junction, where magnetic field distribution exhibits ribbon-like structures. This is a clear manifestation of time-reversal symmetry broken superconducting states in multi-component superconductors.

\vspace{3mm}
\noindent {\it Acknowledgements --} The authors are
grateful to Q. F. Liang, Z. Wang and Y. Takahashi for fruitful discussions. This work was supported by the WPI initiative on Materials Nanoarchitectonics, and by the Grant-in-Aid for Scientific Research (No. 25400385), the Ministry of Education, Culture, Sports, Science and Technology of Japan.

%

\end{document}